\documentclass[aps,endfloat,floatfix,showpacs]{revtex4-1}
\newcommand{\be}{\begin{equation}}
\newcommand{\ee}{\end{equation}}
\newcommand{\bea}{\begin{eqnarray}}
\newcommand{\eea}{\end{eqnarray}}

\usepackage{graphicx}
\usepackage{color}
\usepackage{amsmath}
\usepackage{mathtools}
\usepackage{amsfonts}
\usepackage{bm}

\begin{document}

\title{Fitting ultracold resonances without a fit}
\author{A. Simoni}
\affiliation{
Univ Rennes, CNRS, IPR (Institut de Physique de Rennes)-UMR 6251, F-35000 Rennes, France
}

\date{\today}

\begin{abstract}

We present a numerical procedure allowing one to extract Feshbach resonance
parameters from numerical calculations without relying on approximate
fitting procedures. Our approach is based on a simple decomposition of the
reactance matrix in terms of poles and residual background contribution,
and can be applied to the general situation of inelastic overlapping
resonances. A simple lineshape for overlapping inelastic
resonances, equivalent to known results in the particular cases of isolated and
overlapping elastic features, is also rigorously derived.

\end{abstract}

\maketitle


\section{Introduction}

Feshbach resonances are an ubiquitous phenomenon taking place when
the energy of colliding particles is nearly degenerate
with a quasi-bound level of the system. Resonances are traditionally
observed as a function of collision energy or also by varying applied
electromagnetic fields. In ultracold gases, it has been long realized
that resonances occurring at extremely low energies can be induced by
tuning an external magnetic field \cite{Tiesinga1993}, allowing one
to vary in a controlled way the effective interatomic interaction. The
possibility to control the interaction using resonances is at the hearth
of a series of groundbreaking experiments in the field of ultracold gases;
See {\it e.g.} \cite{Bloch2008}.

Feshbach resonances have been studied both theoretically
and experimentally in a number of ultracold atomic systems
\cite{Chin2010}. Theory usually conveys information extracted from
numerical models in a small number of lineshape parameters, namely position, width
and the non-resonant (or background) contribution, extracted by fitting the numerical
result to an analytical form. A celebrated example is the case
of an isolated elastic resonance \cite{Tiesinga1993} where the $s$-wave scattering length
diverges both positively and negatively as a function of the magnetic
field $B$ across the center $B_0$ :
\be
\label{eq_Fesh_one}
a(B)= a_{\rm bg} \left( 1 - \dfrac{\Delta}{B-B_0}\right) .
\ee
Here $\Delta$ represents the magnetic width and the background
 scattering length $a_{\rm bg}$ is the off-resonance value of $a(B)$. 

A more general lineshape describing a pair and an arbitrary number
of elastic overlapping resonances has been derived using formal scattering theory and
quantum defect analysis in the Supplemental Material of \cite{2013-SR-PRL-053202} and in \cite{Jachymski2013}, respectively.
In this case, the scattering length is still a purely real quantity
showing a pair or a series of nearby divergences.

Isolated resonances in the presence of inelastic processes have
been discussed in \cite{Kempen2002} and studied in greater detail in \cite{Hutson2007,2011-DN-EPJD-55}.
The main influence of inelastic loss channels is to make the scattering
length a complex quantity and to suppress the pole
in its real part, that no longer diverges \cite{Ciurylo2005,Hutson2007}.
In the presence of inelastic processes both the lineshape and the
extraction of related resonance parameters become more involved. A
relatively elaborate fitting strategy based on a three-points iterative
procedure has been developed in \cite{Frye2017}. 
However, like any other approach
relying on fitting the lineshape to an analytical expression, the latter
procedure presents few intrinsic drawbacks. 

First, the precise value of the fitting parameters depends on the chosen
fitting interval and bears an error bar.
Second, the analytical lineshape can be a poor approximation to the
fitted data, even to the extent that the fitting procedures may not
converge~\cite{Frye2017}. Even if the least square
iterations do converge, the physical meaning of the fitting parameters will
become unclear if the analytical lineshape poorly describes the data. For
instance, if the background scattering length presents an unusually rapid
variation, defining such quantity as the off-resonance value of
$a(B)$ is not a well defined prescription.
Finally, the analytical lineshape for overlapping resonances in the
presence of inelastic processes is not yet available. This situation
which indeed occurs frequently for collisions in excited
states~\cite{Frye2019} is still missing a systematic procedure allowing one
to summarize the numerical data in few theoretical parameters. A simple approach 
consisting in presenting graphics of the numerical results
without accompanying tables of resonance parameters is not fully
satisfactory, since reading  quantitative information
from a plot without having access to the rough data can be difficult or impossible.

In this paper, we present a formalism designed to overcome all of the
above difficulties. Our approach allows one to define and extract in an algorithmically
well defined way resonance parameters even in cases where the
lineshape cannot be modeled by a simple expression. For the sake of
better physical insight and clarity, our discussion is based on Feshbach
resonance theory \cite{Feshbach1967}. However, at a fundamental level the present
strategy is based on a general pole
approximation that could be taken as starting point in a similar fashion as
scattering resonances can be defined as scattering-matrix 
poles \cite{Taylor2006}. 
Special attention is granted to stable ways to implement
such pole decomposition, a numerically non trivial task in the presence of naturally diverging 
quantities such as the scattering length on resonance.
As a by-product of the formalism, we also demonstrate that 
the complex scattering length can always be parametrized in a simple and intuitive way in terms of 
complex magnetic field locations and magnetic widths.

The paper is organized as follows. In section~\ref{formalism} we introduce
the formalism building on Feshbach resonance theory.

Section~\ref{comparison} proves that the present approach reduces under
suitable conditions to results previously appeared in the literature.

In section~\ref{applications} we present an application to two situations
of physical interest for $^{39}$K, the case of overlapping and inelastic
$s$-wave resonances and of finite-energy $p$-wave resonances.

A short conclusion ends this work.

\section{Formalism}
\label{formalism}

\subsection{Basic ideas for single channel scattering}
The main idea can be illustrated with reference to the simple single
channel resonant expression of the $s$-wave scattering length in the presence of an isolated resonance. Here we
consider the most common case that the scattering length is tuned via
an external magnetic field $B$, but the formalism applies equally well to a
general external fields or variation of any physical parameter in the
Hamiltonian capable of inducing resonances. In the case of a magnetic
field, recalling \eqref{eq_Fesh_one}, one can write :
\be
a(B)= a_{\rm bg} \left( 1 - \dfrac{\Delta}{B-B_0}\right) \equiv a_{\rm bg} + \dfrac{p}{B-B_0}  
\ee
where we have defined $p=-a_{\rm bg} \Delta$ the {\it
field-independent} pole strength. Any variation of $a$ with $B$ beyond the
pole contribution is included in the background term, which will therefore
be considered a function of $B$. 

The pole strength can be extracted from a theoretical calculation of $a$ as
\be
p = {\lim}_{B \to B_0} \left[ a(B) (B - B_0 ) \right] .
\ee 
It will be shown below how this limit can be performed in a numerically
stable way; See equation~\eqref{pole}.  Once the $p$ parameter is determined, one
can compute the {\it on-resonance} value of the background scattering
length according to 
\be
\label{abg_simple}
a_{\rm bg}(B_0)={\lim}_{B \to B_0}  \left[ a(B) - \dfrac {p}{B-B_0} \right] .
\ee
Again, this limit can be computed in a stable way; See \eqref{eq_aloc}.  
We stress that the usual interpretation of the background scattering length
as the limit value of $a$ far away from resonance may become ambiguous
in cases where $a_{\rm bg}$ is not strictly constant or, even worse,
is rapidly varying. On the converse, the quantity \eqref{abg_simple}
is well defined in any case and denotes simply the value of $a(B)$
{\it at the resonance location} after subtraction of the divergent part.
We now generalize these ideas to multichannel scattering.
\subsection{Pole decomposition for multichannel scattering}
Les us consider a general scattering problem with $N$ channel, $N_c$ closed channels, and $N_{op}=
( N-N_c )$ open channels. We will further partition the $N_{op}$
open channels into $N_o$ degenerate (also termed in scattering theory super-elastic) channels with energy
equal to the incoming channel, and $N_{\bar o}$ inelastic channels with
energy release (or absorption). We use Feshbach theory, that formally
partitions channel space into open and closed channel subspaces $P$ and $Q$ \cite{Feshbach1958}. 
Scattering in the $P$ subspace is supposed to be
known, and the influence of the $Q$ subspace is taken into account through
the introduction of an effective Hamiltonian. The nature of $P$ and $Q$
subspaces depends on the specific problem at hand
and for general formal manipulations does not even need to be specified \cite{1983-WD-PRA-2777}. 
A resonance occurs when the energy of bound states 
in $Q$ subspace is quasi-degenerate with the energy of the colliding particles. 
Under such conditions the influence of the $Q$ subspace on the total scattering amplitude 
becomes dramatic.

Feshbach formalism is most often applied to the scattering
matrix~\cite{Feshbach1958}, but also provides a partition of the reactance
matrix~\cite{Feshbach1967}. 
Working with the reactance matrix is convenient
since the resulting scattering matrix is automatically unitary, algebra is real, and the number
of fitting parameters is minimal~\cite{Feshbach1967}.  
According to Feshbach theory, the reactance
matrix at total energy $E$ can be written
\bea
\label{eq_kmat_Fesh}
{\mathbf K} &=& \left[ \sin({\bm \xi}_{\rm bg}) + \cos({\bm \xi}_{\rm bg})  {\mathbf K}_{\rm
res}  \right]  \left[ \cos({\bm \xi}_{\rm bg}) - \sin({\bm \xi}_{\rm bg}) {\mathbf K}_{\rm
res}   \right]^{-1} \label{kmatgen:1}  \\
&=& \cos({\bm \xi_{\rm bg}}) \left[ \tan({\bm \xi}_{\rm bg}) + {\mathbf
K}_{\rm res}  \right]  \left[ {\bm 1}_P - \tan({\bm \xi}_{\rm bg}) {\mathbf K}_{\rm
res}   \right]^{-1} \cos({\bm \xi_{\rm bg}})^{-1} \label{kmatgen:2} 
\eea
where ${\bm 1}_P$ is the identity in $P$ space and the resonant part of the reactance matrix has the form
\be
{\mathbf K}_{\rm res}= {\mathbf y} \left( E {\bm 1}_Q - {\mathbf E}_b - {\mathbf V}  \right)^{-1} {\mathbf y}^t .
\ee
Here $\mathbf y$ is the $N_{op} \times N_c$ matrix formed by half predissociation
amplitudes and the operator $\left( E{\bm 1}_Q - {\mathbf E}_b - {\mathbf V}
\right)$ acts in the closed channel subspace, the $Q$-space of Feshbach
theory. Zero energy is taken at the energy of the separated atoms at
the given value of the applied field. The diagonal matrix ${\mathbf E}_b$
represents the energy levels of the bare molecular states defining the
$Q$-space. We will make the usual assumption that the position of such
levels depends linearly on the applied field ${\mathbf E}_b=-{\bm \mu} (
B{\bm 1}_Q - {\mathbf B}_r ) $ with $\bm \mu$ a diagonal matrix whose entries
$\mu_\alpha$ are the differential magnetic moment of state $\alpha$ with
respect to the one of two separated atoms in the initial channel. The
entries $B_{r,\alpha}$ of the diagonal matrix ${\mathbf B}_r$ represent the
bare resonance magnetic field, {\it i.e} the field at which level $\alpha$
crosses the incoming channel dissociation threshold.
The coupling $\mathbf V$ contains in general an energy-independent direct
coupling $\mathbf U$ between molecular states as well as indirect coupling
through the continuum via the level shift operator ${\bm \delta}^E$ at
energy $E$~\cite{CohenTannoudji1992,Simoni2002}. At this stage it is
not necessary to distinguish the two contributions, only the Hermitic
nature of $\mathbf V$ being of importance.

To proceed further with the derivation, inspired by the single-channel case, we factor out the
magnetic moment in the denominator \cite{Tiesinga1993}. To this aim, we define $\tilde{\mathbf y}={\bm
\mu}^{-1/2} {\mathbf y}$ and $ {\bm {\mathcal B}} =  {\mathbf B}_r 
+ {\bm \mu}^{-1/2} ({\mathbf V}-E{\bm 1}_Q) {\bm \mu}^{-1/2} $. The resonant reactance
matrix is expressed in terms of these redefined quantities as :
\be
\label{kresredef}
 {\mathbf K}_{\rm res}= \tilde{\mathbf y} \left( B{\bm 1}_Q - {\bm {\mathcal B}}  \right)^{-1} \tilde{\mathbf y}^t .
\ee
If the magnetic moments all share the same sign the matrix ${\bm
\mu}^{1/2}$ will be either purely real or purely imaginary. In this
case, by its definition $\bm {\mathcal B}$ will be real and symmetric,
thus with real eigenvalues. If the magnetic moments present both negative
or positive values the matrix $\bm  {\mathcal B}$ will in general be
complex, yet still symmetric.

We now insert \eqref{kresredef} into~\eqref{kmatgen:2}. 
After some algebra, see proof in Appendix~\ref{appendix_1}, one can recast the $K$ matrix in the form
\be
\label{kmatgenB}
{\mathbf K} =  \tan({\bm \xi}_{\rm bg}) + \cos({\bm \xi}_{\rm bg})^{-1} \tilde{\mathbf y} (B{\bm 1}_Q-\hat{\bm{\mathcal B}})^{-1}    \tilde{\mathbf y}^t  \cos({\bm \xi}_{\rm bg})^{-1}
\ee
where the shifted resonance position matrix $\hat{\bm{\mathcal B}}$
is defined as $\hat{\bm{\mathcal B}}=\bm{\mathcal B} + \tilde{\mathbf
y}^t  \tan({\bm \xi}_{\rm bg}) \tilde{\mathbf y}  $. If there are no
inelastic channels and if threshold collisions are considered ($E\to 0$), all
elements of $\tan({\bm \xi}_{\rm bg})$ as well as of the predissociation
amplitude $\mathbf y$ tend to zero, hence $\hat{\bm{\mathcal B}} \to
\bm{\mathcal B}$.
The $\hat{\bm{\mathcal B}}$ operator is still complex symmetric, is real if ${\bm{\mathcal B}}$ is real,
but unlike ${\bm{\mathcal B}}$ in general it
varies with magnetic field because of the background phase ${\bm \xi}_{\rm bg}$ appearing in its definition.
It can therefore be brought a diagonal form $\hat{\mathbf{b}}$ through a 
$B$-dependent complex orthogonal
transformation $\mathbf C$ such that $ \hat{\bm{\mathcal B}} = {\mathbf
C} \hat{\mathbf{b}} {\mathbf C}^t$ and ${\mathbf C}^t {\mathbf C}
= {\mathbf C} {\mathbf C}^t = {\mathbf 1}$~\cite{Wilkinson1965}.
With this transformation, equation~\eqref{kmatgenB} can be rewritten :
\be
\label{kmatgenB_a}
{\mathbf K} =  \tan({\bm \xi}_{\rm bg}) +  \hat{\mathbf y} (B{\bm 1}_Q-\hat{\mathbf{b}})^{-1}    
\hat{\mathbf y}^t  = \tan({\bm \xi}_{\rm bg}) + \sum_\alpha  
\frac{\hat{\mathbf y}_\alpha \hat{\mathbf y}^t_\alpha}{B-{\hat b}_\alpha}
\ee
in terms of a redefined predissociation amplitude $\hat{\mathbf y}=\cos({\bm \xi}_{\rm bg})^{-1}  \tilde{\mathbf y} {\mathbf C}^{t}$, which also depends on $B$.

The reactance matrix clearly diverges at the solutions, say $b_\alpha$,
of the nonlinear equation $B-{\hat b}_\alpha(B)= 0$.
The following manipulations will turn out to be simpler if one isolates 
pole contributions with a field-independent pole residue.
To this aim, let us define the constant amplitude vector $\hat{\mathbf y}^{(0)}_\alpha=\hat{\mathbf y}_\alpha(b_\alpha)$ 
and rewrite \eqref{kmatgenB_a} 
\be
\label{kmatgenB_b}
{\mathbf K} =  \tan(\tilde{\bm \xi}_{\rm bg}) +  \hat{\mathbf y}^{(0)} (B{\bm 1}_Q-{\mathbf{b}})^{-1}    
\hat{\mathbf y}^{(0)t}  = \tan(\tilde {\bm \xi}_{\rm bg}) + \sum_\alpha  
\frac{\hat{\mathbf y}^{(0)}_\alpha \hat{\mathbf y}^{(0)t}_\alpha}{B-b_\alpha} .
\ee
where the modified background
\be
\tan(\tilde {\bm \xi}_{\rm bg}) = \tan({\bm \xi}_{\rm bg})
-\hat{\mathbf y}^{(0)} (B{\bm 1}_Q-{\mathbf{b}})^{-1}
\hat{\mathbf y}^{(0)t} + 
\hat{\mathbf y} (B{\bm 1}_Q-\hat{\mathbf{b}})^{-1}
\hat{\mathbf y}^t  .
\ee
is still regular at the poles and reduces 
to $\tan({\bm \xi}_{\rm bg})$ if ${\bm \xi}_{\rm bg}$ is strictly 
constant.

According to the remark below \eqref{kresredef} only real values of $b_\alpha$ are possible if all entries in matrix ${\bm \mu}$ have the same sign.
If this is not the case, it can be shown that 
solutions $b_\alpha$ are either real or occur in pairs of complex conjugate quantities.
On a physical ground, such complex solutions arise when a pair of molecular levels with opposite slopes avoid each other
exactly at the collision threshold, such that neither level crosses the latter. 
For purely elastic scattering,
this unlikely case will result in
a scattering length that may have a rapid variation, but will not diverge as $B$ varies
on the real axis. We will not consider this situation as a resonance condition and will henceforth assume that $b_\alpha$ is always real, with the 
implicit assumption that the contribution from
complex solutions $b_\alpha$, if any, is included in the background.
 
Since we are mainly interested in scattering near a collision threshold, it is
convenient to factor out the low-energy limit behavior of the reactance
matrix for channels with orbital angular momentum $\ell$ as follows \cite{RHD-1961-RMP-471}:
\be
\label{eq_norm}
\bar{\mathbf K} = {\mathbf q}^{-(\ell + 1/2)} {\mathbf K}  {\mathbf q}^{-(\ell + 1/2)}  .
\ee
The diagonal normalization matrix ${\mathbf q}^{-(\ell + 1/2)}$ has by
definition elements $q_i^{-(\ell_i + 1/2)}$ for all channels degenerate
with the incoming one, and to 1 for inelastic ones. The resulting
normalized $K$-matrix is thus finite at threshold and is the main quantity
needed to compute zero-energy scattering lengths. Performing such
channel normalization and introducing the background scattering length
matrix ${\mathbf A}={\mathbf q}^{-(\ell + 1/2)} \tan(\tilde{\bm \xi}_{\rm
bg})  {\mathbf q}^{-(\ell + 1/2)}$ as well as the resonance amplitude
$\bar{\mathbf y}={\mathbf q}^{-(\ell + 1/2)} \hat{\mathbf y}^{(0)}$,
equation~\eqref{kmatgenB_b} transforms into
\be
\label{kmatgenB_c}
\bar{\mathbf K} =  {\mathbf A} +  \bar{\mathbf y} (B{\bm 1}_Q-{\mathbf{b}})^{-1}    
\bar{\mathbf y}^t  = {\mathbf A}  + \sum_\alpha  
\frac{\bar{\mathbf y}_\alpha \bar{\mathbf y}^t_\alpha}{B-b_\alpha} .
\ee
Note that according to the Wigner laws \cite{Wigner1948}, by construction both ${\mathbf
A}$ and $\bar{\mathbf y}$ are finite at threshold.

With the assumption above, the matrix $\bar{\mathbf K}$ diverges at the real
magnetic field locations $b_\alpha$. This should be contrasted with the
behavior of the complex scattering length, that even in the presence of
a pole in $\bar{\mathbf K}$ on the real axis can in some cases present 
barely no structure. 
In order
to locate the poles numerically, we find it more convenient to consider
the inverse reactance matrix ${\mathbf M}={\bar{\mathbf K}}^{-1}$ and
to search for the zeros in the determinant $\det{\mathbf M}$. We first
bracket zeros by a coarse search, then quickly refine their position
using Ridder's method. Such nonlinear root-finding algorithm has as
main advantages a superlinear convergence and that of keeping the root
bracketed \cite{Press1996}.

We now show that the residue as well as the background contribution {\it
at the pole position} (recall that ${\mathbf A}$ will, in general, have
a dependence on $B$) can be accurately determined.
Concerning the residue, by definition :
\be
\bar{\mathbf y}_\alpha \bar{\mathbf y}^t_\alpha= 
{\lim}_{B\to b_\alpha}  (B-b_\alpha)  \bar{\mathbf K}(B) .
\ee
In order to evaluate the indeterminate $0 \cdot \infty$ product, we write
$\bar{\mathbf K}={\mathbf M}^{-1}= \text{adj}{~\mathbf M}/\det{\mathbf
M}$ and use l'H\^{o}pital rule. Keeping into account continuity of the
adjugate, one promptly obtains 
\be 
\label{pole}
\bar{\mathbf y}_\alpha \bar{\mathbf y}^t_\alpha =
\dfrac{\text{adj}{~\mathbf M(b_\alpha)}}{\det^\prime{\mathbf M}(b_\alpha)} .
\ee
Equation \eqref{pole} gives a numerically stable way to evaluate the residue in
terms of $\mathbf M$. The needed derivatives of the determinant can be
evaluated using Jacobi's formula :
\be
\label{eq_ddet}
{\det}^\prime{\mathbf M}(B) =\text{tr} \left( \text{adj}({\mathbf M}(B))  {\mathbf M}^\prime     \right)
\ee

The local background at the pole is defined by subtracting the divergent contribution from the global reactance matrix :
\be
{\mathbf A}_{\text {loc}}(b_\alpha)  = {\lim}_{B \to b_\alpha}
\left( \bar{\mathbf K}(B) - \frac{\bar{\mathbf y}_\alpha \bar{\mathbf
y}^t_\alpha}{B-b_\alpha} \right)  .
\ee
Note that ${\mathbf A}_{\text {loc}}$ contains contributions from the
global background ${\mathbf A}$ as well as from all resonances
but the $\alpha$-th one. The lhs presents itself in a $\infty - \infty$
form. In order to evaluate the limit, we write again $\bar{\mathbf K}$
in terms of the adjugate and determinant of $\mathbf M$. After bringing
to common denominator and Taylor expanding to second order one obtains
\be
\label{eq_aloc}
{\mathbf A}_{\text {loc}}(b_\alpha)  = \dfrac{2\det^\prime({\mathbf M}(b_\alpha)) 
\text{adj}^\prime({\mathbf M(b_\alpha)}) - \det^{\prime \prime}{\mathbf M}(b_\alpha) 
\text{adj}({\mathbf M(b_\alpha)} }
{2 \left[ \det^\prime{\mathbf M}(b_\alpha) \right]^2} .
\ee
One proceeds similarly for the first derivative, which will also be needed. One starts from the definition 
\be
{\mathbf A}_{\text {loc}}^\prime(b_\alpha)  = {\lim}_{b \to b_\alpha}
\left( \bar{\mathbf K}^\prime(B) + \frac{\bar{\mathbf y}_\alpha \bar{\mathbf
y}^t_\alpha}{(B-b_\alpha)^2} \right)  
\ee
and evaluates the indeterminate limit using Taylor expansion to get
\be
\label{eq_dAbgl}
{\mathbf A}_{\text {loc}}^\prime(b_\alpha)  = 
\dfrac{\splitfrac{ 2 (\det^{\prime \prime}{\mathbf M}(b_\alpha))^2 \text{adj}({\mathbf M(b_\alpha)}     -2\det^\prime({\mathbf M}(b_\alpha)) \det^{\prime \prime}({\mathbf M}(b_\alpha)) \text{adj}^\prime({\mathbf M(b_\alpha)})
+2 (\det^\prime{\mathbf M}(b_\alpha))^2 \text{adj}^{\prime \prime}({\mathbf M(b_\alpha)})}
{- \det^\prime({\mathbf M}(b_\alpha)) \det^{\prime \prime \prime}({\mathbf M}(b_\alpha)) \text{adj}({\mathbf M(b_\alpha)})
-\det^\prime({\mathbf M}(b_\alpha)) \det^{\prime \prime }({\mathbf M}(b_\alpha)) \text{adj}^{\prime \prime}({\mathbf M(b_\alpha)})
}}
{2 (\det^\prime{\mathbf M}(b_\alpha))^3  }  .
\ee
The required derivatives of the adjugate can be evaluated keeping into account that elements of the adjugate
are in turn determinants of minors, and using again~\eqref{eq_ddet}. 
Higher order derivatives of a determinant (and thus of the adjugate) follow by successive differentiation of~\eqref{eq_ddet}.
For instance, the second derivative can be calculated as
\be
 {\det}^{\prime \prime}{\mathbf M}(B) =\text{tr} \left( \text{adj}^\prime({\mathbf M}(B)) 
{\mathbf M}^\prime  +\text{adj}({\mathbf M}(B))  {\mathbf M}^{\prime \prime}   \right) .
\ee


Once the local background matrix (which includes contributions from nearby resonances) has been determined, 
the usually smooth global background and its derivative are respectively given by
\be
{\mathbf A}(b_\alpha)={\mathbf A}_{\text {loc}}(b_\alpha)-
\sum_{\beta \neq \alpha}
\frac{\bar{\mathbf y}_\beta \bar{\mathbf y}^t_\beta}{b_\alpha-b_\beta} .
\ee
and
\be
{\mathbf A}^\prime(b_\alpha)=  {\mathbf A}_{\text {loc}}^\prime(b_\alpha)+
\sum_{\beta \neq \alpha}
\frac{\bar{\mathbf y}_\beta \bar{\mathbf y}^t_\beta}{(b_\alpha-b_\beta)^2} .
\ee

In practice, it is clear that in these expansions one will not need
to include all poles of $\mathbf K$. One rather focuses on resonances
occurring in a predefined magnetic field region and includes all poles
such that the addition of additional ones has negligible influence on
the parameters of the resonances of interest. In simpler terms, only
resonances significantly overlapping with the ones under investigation must
be taken into account. Is is also clear that without
a stable and accurate algorithm to compute the $\mathbf M$ matrix and
its derivatives the present formalism would be of little practical use.
Our numerical approach for computing such derivatives is the subject of
the following subsection.

\subsection{Computing the derivatives}

On a general ground, a basis representation of the wavefunction
is particularly suited to compute derivatives of any order, since the
latter can then be obtained in an essentially analytical fashion.
In fact, while propagation algorithms also can be adapted for obtaining derivatives (see {\it e.g.} \cite{VA-2005-JCP-054314}),
algorithmic complexity and computational cost both increase with the derivative order.
In this work we exploit the pseudospectral representation termed spectral element method,
a spatial grid representation presented for instance in \cite{2017-AS-JCP-244106} and references therein.
In the present work we impose asymptotic $R$-matrix boundary conditions, but one could equally 
well work with log-derivative or scattering boundary conditions.
In the discrete spatial grid representation, the time-independent multichannel Schr\"odinger equation reads : 
\be
\label{eq_var_schroedinger}
[ E{\mathbf 1}-{\mathbf H}(B) ] {\mathbf \Psi}(B) = {\mathbf c}
\ee
where the constant vector $\mathbf c$ on the rhs equal to zero everywhere but at last grid point $r_{\rm max}$, 
where it equals the unit matrix of dimension the total number of channels~\cite{2017-AS-JCP-244106}.
Elements of the Hamiltonian matrix in the spectral element representation can be found in \cite{2017-AS-JCP-244106}.
By definition, the multichannel wavefunction $\mathbf \Psi$ at $r_{\rm max}$ equals the $R$-matrix $\mathbf R$.

Once ${\mathbf \Psi}$ has been determined from equation \eqref{eq_var_schroedinger},
first derivative of the matrix wavefunction is obtained from differentiating~\eqref{eq_var_schroedinger} with
respect to parameter $B$, and solving the resulting linear system:
\be
{\mathbf H}(B) \frac{\partial \mathbf \Psi}{\partial B}    +  \frac{\partial \mathbf H}{\partial B} {\mathbf \Psi}(B) 
=0 .
\ee
Again, the derivative of the $R$-matrix with respect to $B$ corresponds to the computed ${\partial \mathbf \Psi}/{\partial B}$ at $r_{\rm max}$.

The $K$-matrix is ordinarily computed by imposing the asymptotic form of the 
wavefunction ${ \mathbf \Psi } \sim {\mathbf f } - {\mathbf g} {\mathbf K}$ in terms of regular
and irregular solutions $\mathbf f$ and $\mathbf g$ for the free particle problem.
Since in our algorithm $\mathbf K$ is not needed, we rather determine directly its inverse $\mathbf M$
by matching to ${ \mathbf \Psi } \sim {\mathbf f } {\mathbf M}- {\mathbf g}$.
The linear matching equations for $\mathbf M$ simply follow from equation~(25) of \cite{2017-AS-JCP-244106} and read : 
\be
\label{eq_match}
({\mathbf f} -{\mathbf R} {\mathbf f}^\prime )  {\mathbf M} = ({\mathbf g} -{\mathbf R} {\mathbf g}^\prime )
\ee
the prime denotes the spatial derivative and all spatial arguments are tacitly taken at last point
of the discretization grid.
The first derivative of ${\mathbf M}$ with respect to the field is solution to the linear system obtained taking the derivative of \eqref{eq_match}
\be
({\mathbf f} -{\mathbf R} {\mathbf f}^\prime )  \frac{\partial \mathbf M}{\partial B}
+ \frac{\partial ({\mathbf f} -{\mathbf R} {\mathbf f}^\prime )}{\partial B} {\mathbf M}
= \frac{\partial \mathbf g}{\partial B} -{\mathbf R} \frac{\partial {\mathbf g}^\prime}{\partial B}
- \frac{\partial \mathbf R}{\partial B} {\mathbf g}^\prime
\ee
in which $\mathbf M$, $\mathbf R$, and $\dfrac{\partial \mathbf R}{\partial B}$ are known.
Derivatives of the reference functions with respect to the field are transformed into derivatives with
respect to energy using chain rule and the known behavior of asymptotic channel energies with respect to $B$. 
The latter are calculated using a stable formula for both open
and closed channels from \cite{VA-2005-JCP-054314}, equations (11) and (12).

Formal generalization to the $n$-th derivative of $\mathbf \Psi$ is straightforward.
One differentiates $n$ times the Schr\"odinger equation, and solves for the highest order derivatives of the
wavefunction 
\be
  \sum_k  \binom{n}{k} \frac{\partial^k \mathbf H}{\partial B^k}  \frac{\partial^{n-k} \mathbf \Psi}{\partial B^{n-k}} =0
\ee
knowing the derivatives of $\mathbf \Psi$ up to order $n-1$.
Similarly, to determine $\partial^n{\mathbf M} / \partial B^n$  one differentiates $n$ times 
the matching condition \eqref{eq_match}
\be
 \sum_k  \binom{n}{k} \frac{\partial^k {(\mathbf f -{\mathbf R} {\mathbf f}^\prime)}}{\partial B^k}  \frac{\partial^{n-k} \mathbf  M}{\partial B^{n-k}}   =  \frac{\partial^{n} \mathbf  g}{\partial B^{n}} - 
\sum_k  \binom{n}{k} \frac{\partial^k {\mathbf R}}{\partial B^k}  \frac{\partial^{n-k} \mathbf  g^\prime}{\partial B^{n-k}}
\ee
and solves for the highest order derivative knowing the derivatives of $\mathbf M$ up to order
$n-1$ as well as the derivatives of $\mathbf R$ up to order $n$. Higher order derivatives of reference functions
with respect to $B$ are obtained by successive differentiation of equations (11) and (12) in \cite{VA-2005-JCP-054314}.

\subsection{Observables : the energy-dependent scattering length}

We now turn to the calculation of observable quantities. The scattering cross-sections at arbitrary energy can be expressed exactly 
as a function of the so-called energy-dependent scattering length~\cite{Bolda2003}.
Its definition for a single channel in terms of the diagonal scattering matrix element $S_{jj}$ :
\be
a(B)= q_o^{-(2 \ell_i + 1)} \frac{1 - S_{jj}}{i(1 + S_{jj})}
\ee
where $q_o$ is the channel wavevector.

We consider here the general situation of a 
number $N_{o}$ of degenerate channels occurring
for example when several partial waves contribute to the scattering process or in the presence
of magnetic Zeeman sublevels in a vanishing magnetic fields.
Hence we define a square complex scattering length matrix
\be
\label{eq_asc}
{\mathbf a}(B)=  {{\mathbf q_o}^{- (\ell_i + 1/2)}} \frac{{\mathbf 1}_{ o  o}  - {\mathbf S}_{ o  o}}{i\left( {\mathbf 1}_{ oo} + {\mathbf S}_{ o  o} \right) } {{\mathbf q_o}^{- (\ell_i + 1/2)}} .
\ee
The off-diagonal elements of $\mathbf a$ are particularly relevant 
to parametrize anisotropic pseudopotentials~\cite{2003-AD-PRA-033607} or
in the context of spinor Bose-Einstein condensates \cite{2012-YK-PR-243}.

We now write the relevant square matrices ${\mathbf X}={\mathbf K},{\mathbf A},{\mathbf 1}$ as well as the generally
non-square predissociation amplitudes $\mathbf{y}$ in blocked form for open degenerate and non-degenerate channels :
\be 
{\mathbf X}=
\left[
\begin{array}{c|c}
{\mathbf X}_{oo} & {\mathbf X}_{o {\bar o}} \\
\hline
{\mathbf X}_{{\bar o} o} & {\mathbf X}_{{\bar o} {\bar o}}
\end{array}
\right]  
~ , ~     
{\mathbf {\bar y}}=
\left[
\begin{array}{c}
{\mathbf {\bar y}}_{o}  \\
\hline
{\mathbf {\bar y}}_{{\bar o} } 
\end{array}
\right] .
\ee
With this decomposition and applying the projection technique of \cite{RHD-1961-RMP-471} 
one can express the middle matrix ratio on the rhs of \eqref{eq_asc} as a function of the blocks of the reactance matrix
\be
\frac{{\mathbf 1}_{ o  o}  - {\mathbf S}_{ o  o}}{i( {\mathbf 1}_{ oo} + {\mathbf S}_{ o  o}) }
= - {\mathbf K}_{ o  o} -i  {\mathbf K}_{ o  {\bar o}} \left(  {\mathbf 1}_{ {\bar o}{\bar o}} -i  {\mathbf K}_{ {\bar o} {\bar  o} }    \right)^{-1}  {\mathbf K}_{ {\bar o}  o}
\ee
Transforming to the normalized quantities according to \eqref{eq_norm}
\be
{\mathbf a}(B)
= - \bar{ \mathbf K}_{ o  o} - i \bar{\mathbf K}_{ o  {\bar o}} \left(  {\mathbf 1}_{ {\bar o}{\bar o}} -i \bar{\mathbf K}_{ {\bar o} {\bar  o} }    \right)^{-1} \bar{\mathbf K}_{ {\bar o}  o}
\ee
and with the help of \eqref{kmatgenB_c} we obtain :
\bea
\label{eq_a_en_dep}
   {\mathbf a}(B) &=& - {\mathbf A}_{oo}  - \bar{\mathbf y}_{o} \left( B {\mathbf 1}_Q - {\mathbf b} \right)^{-1} \bar{\mathbf y}^t_{o} \nonumber \\ 
 &-& i\left(  {\mathbf A}_{o\bar o} +  \bar{\mathbf y}_{o} \left( B {\mathbf 1}_Q - {\mathbf b} \right)^{-1} \bar{\mathbf y}^t_{\bar o}     \right)
\left(  {\mathbf 1}_{\bar o \bar o} -i {\mathbf A}_{\bar o \bar o} -i \bar{\mathbf y}_{\bar o} \left( B {\mathbf 1}_Q - {\mathbf b} \right)^{-1} \bar{\mathbf y}^t_{\bar o}  \right)^{-1}   
\left( {\mathbf A}_{\bar o o} +  \bar{\mathbf y}_{o}  \left( B{ \mathbf 1}_Q - {\mathbf b} \right)^{-1} \bar{\mathbf y}^t_{\bar o}     \right).  
\eea
Making use of the Woodbury matrix identity \cite{wiki:001} one can further write :
\bea
\label{eq_wood}
&& \left(  {\mathbf 1}_{\bar o \bar o} - i {\mathbf A}_{\bar o \bar o} 
- i \bar{\mathbf y}_{\bar o} \left( B {\mathbf 1}_Q 
- {\mathbf b} \right)^{-1} \bar{\mathbf y}^t_{\bar o}  \right)^{-1}  
= \left(  {\mathbf 1}_{\bar o \bar o} - i {\mathbf A}_{\bar o \bar o}  \right)^{-1}  \nonumber \\
&+& i \left(  {\mathbf 1}_{\bar o \bar o} -i  {\mathbf A}_{\bar o \bar o}  \right)^{-1}  \bar{\mathbf y}_{\bar o} 
\left( B{ \mathbf 1}_Q - {\mathbf b} -i \bar{\mathbf y}^t_{\bar o} \left( {\mathbf 1}_{\bar o \bar o} 
- {\mathbf A}_{\bar o \bar o}   \right)^{-1} \bar{\mathbf y}_{\bar o}  \right)^{-1}
\bar{\mathbf y}^t_{\bar o}  \left(  {\mathbf 1}_{\bar o \bar o} -i {\mathbf A}_{\bar o \bar o}  \right)^{-1}.
\eea

Substituting \eqref{eq_wood} into \eqref{eq_a_en_dep}, defining
a new predissociation matrix ${\mathbf u}_{\bar o}= 
\left(  {\mathbf 1}_{\bar o \bar o} -i  {\mathbf A}_{\bar o \bar o}  \right)^{-1}  \bar{\mathbf y}_{\bar o}  $ and 
the complex resonance position 
${\mathbf B}_c={\mathbf b} +i \bar{\mathbf y}^t_{\bar o} \left( {\mathbf 1}_{\bar o \bar o} 
- {\mathbf A}_{\bar o \bar o}   \right)^{-1} \bar{\mathbf y}_{\bar o} $,
equation \eqref{eq_a_en_dep} can be recast in the more compact form
\bea
\label{eq_int}
{\mathbf a}(B)  &=& 
- {\mathbf A}_{oo}  - \bar{\mathbf y}_{o} \left( B {\mathbf 1}_Q - {\mathbf b} \right)^{-1} \bar{\mathbf y}^t_{o} \nonumber \\
&-& i  \left( {\mathbf A}_{o \bar o} +  \bar{\mathbf y}_{o}  ( B {\mathbf 1}_Q - {\mathbf b})^{-1} \bar{\mathbf y}^t_{\bar o} \right)
\left(  (  {\mathbf 1}_{\bar o \bar o} -i {\mathbf A}_{\bar o \bar o}  )^{-1}  + i  {\mathbf u}_{\bar o} ( B {\mathbf 1}_{Q} - {\mathbf B}_c)^{-1}   {\mathbf u}^t_{\bar o} \right)
 \left( {\mathbf A}_{\bar o o} +  \bar{\mathbf y}_{\bar o}  ( B {\mathbf 1}_Q - {\mathbf b})^{-1} \bar{\mathbf y}^t_{o} \right) . \nonumber \\ 
\eea
With some additional algebra and the definition $\bar {\mathbf u}_{ o}= i {\mathbf A}_{o{\bar o}} {\mathbf u}_{\bar o} + \bar{\mathbf y}_{ o}$ it is possible to put \eqref{eq_int} in a 
form that clearly exhibits the divergence of the complex scattering length
at the complex eigenvalues of the matrix ${\mathbf B}_c$ :
\bea
\label{eq_c_pole}
{\mathbf a}(B) &=& - {\mathbf A}_{oo} -i {\mathbf A}_{o{\bar o}} \left( {\mathbf 1}_{{\bar o}{\bar o}} -i {\mathbf A}_{{\bar o}{\bar o}} \right)^{-1} {\mathbf A}_{{\bar o} o} - \bar{\mathbf u}_{o} ( B {\mathbf 1}_{Q} - {\mathbf B}_c)^{-1}   \bar{\mathbf u}_{o}^t  .
\eea
It should be noted that the matrix function ${\mathbf B}_c$ and therefore its eigenvalues
$B_{c,i}$ depend on the magnetic field $B$ through the dependence of ${\mathbf B}_c$ on ${\mathbf A}_{{\bar o}{\bar o}}$, which in general vary with $B$.
In order to carry out the pole expansion one can proceed in a similar way as it has been done for the $K$-matrix.
First one writes
\be
 ( B {\mathbf 1}_{Q} - {\mathbf B}_c)^{-1} = \dfrac{\text{adj}{~( B {\mathbf 1}_{Q} - {\mathbf B}_c)}}{\det{( B {\mathbf 1}_{Q} - {\mathbf B}_c)}}
\equiv  \dfrac{{\mathbf C}(B)}{\det{( B {\mathbf 1}_{Q} - {\mathbf B}_c)}}
\ee
where the determinant can be expressed as the product of the eigenvalues of its matrix argument 
\be
 \det{( B {\mathbf 1}_{Q} - {\mathbf B}_c)} = {\prod}_{\alpha} (B - B_{c,\alpha}(B) ) .
\ee
Hence, it is clear that poles occur at complex magnetic field locations $B=b_{c,\alpha}$ such that $b_{c,\alpha} - B_{c,\alpha}(b_{c,\alpha})=0$.
The complex poles $b_{c,\alpha}$ in \eqref{eq_c_pole} can equivalently be defined as the solutions of the nonlinear eigenvalue
problem ${\mathbf B}_{c}(B) | \Omega_\alpha(B)\rangle = B | \Omega_\alpha(B) \rangle $. 
In practice the dependence on $B$ is usually weak and such nonlinear
problem can be solved in a few iterations with excellent starting point
the eigenvalues and eigenvectors of the linear problem with constant
$\mathbf A = \mathbf A (b_\alpha)$.

Using regularity of the adjugate and omitting for notational
simplicity the common energy dependence, the residue at the pole can be extracted as the limit
\bea
\label{eq_polec}
{\mathbf p}_\alpha&=&\mathop{\rm Res}(\mathbf a,b_{c,\alpha}) = {\lim}_{B \to b_{c,\alpha}}  (B-b_{c,\alpha}) {\mathbf a}(B) 
= - \dfrac{\bar{\mathbf u}_{o}(b_{c,\alpha}) \text{adj}{\left( b_{c,\alpha} {\mathbf 1}_{Q} - {\mathbf B}_c(b_{c,\alpha}) \right)}  
\bar{\mathbf u}_{o}^t(b_{c,\alpha})} {(1-B_{c,\alpha}^\prime(b_{c,\alpha}))  \prod_{\beta \neq \alpha} (b_{c,\alpha} - B_{c,\beta}(b_{c,\alpha}) )}
   \nonumber \\
 &=& -\dfrac{\bar{\mathbf u}_{o}(b_{c,\alpha}) {\mathbf C}(b_{c,\alpha})  
\bar{\mathbf u}_{o}^t(b_{c,\alpha})} {(1-B_{c,\alpha}^\prime(b_{c,\alpha})) D_\alpha(b_{c,\alpha})}, 
\eea
where we have defined $ D_\alpha(B) =  \prod_{\beta \neq \alpha} (B - B_{c,\beta}(B) )$.

The local complex background at the pole $b_{c,\alpha}$ is defined by subtracting the contribution of pole $\alpha$ from the full
complex scattering length :
\be
{\mathbf a}_{{\rm bgl} , \alpha} = {\lim}_{B\to b_{c,\alpha}}\left[ {\mathbf a}(B) - \dfrac{{\mathbf p}_\alpha}{B-b_{c,\alpha}} \right]
\ee
Substituting \eqref{eq_polec} for the pole strength, bringing to common denominator and applying 
l'H\^{o}pital's rule twice, one obtains
\bea
\label{eq_abgl_c}
{\mathbf a}_{{\rm bgl} , \alpha} &=&  - {\mathbf A}_{oo}(b_{c,\alpha}) -i {\mathbf A}_{o{\bar o}}(b_{c,\alpha}) \left( {\mathbf 1}_{{\bar o}{\bar o}} 
-i {\mathbf A}_{{\bar o}{\bar o}}(b_{c,\alpha}) \right)^{-1} {\mathbf A}_{{\bar o} o}(b_{c,\alpha})
 - \dfrac{  B_{c,\alpha}^{\prime \prime}(b_{c,\alpha})  \bar{\mathbf u}_{o}(b_{c,\alpha}) {\mathbf C}(b_{c,\alpha})  \bar{\mathbf u}_{o}^t(b_{c,\alpha})}{2 (1 - B_{c,\alpha}^\prime(b_{c,\alpha}))^2 D_\alpha(b_{c,\alpha})}  \nonumber \\
&-& \dfrac{ (\bar{\mathbf u}_{o} {\mathbf C} \bar{\mathbf u}_{o}^t   )^\prime(b_{c,\alpha})}{  (1 - B_{c,\alpha}^\prime(b_{c,\alpha}))  D_\alpha(b_{c,\alpha})}  
+ \dfrac{ D_\alpha^\prime(b_{c,\alpha}) \bar{\mathbf u}_{o}(b_{c,\alpha}) {\mathbf C}(b_{c,\alpha})  
\bar{\mathbf u}_{o}^t(b_{c,\alpha}) }{(1 - B_{c,\alpha}^\prime(b_{c,\alpha})) D_\alpha^2(b_{c,\alpha})  }. 
\eea
This quantity contains contributions from all poles but $\alpha$-th one.
The global complex background at the pole $b_{c,\alpha}$ is defined by subtracting all other pole contributions :
\be
{\mathbf a}_{{\rm bg} , \alpha} = {\mathbf a}_{{\rm bgl} , \alpha}  - \sum_{\beta \neq \alpha} \dfrac{{\mathbf p}_\beta}{b_{c,\alpha}-b_{c,\beta}} 
\ee
resulting in
\bea
\label{eq_abg_c}
{\mathbf a}_{{\rm bg} , \alpha} &=&  - {\mathbf A}_{oo}(b_{c,\alpha}) -i {\mathbf A}_{o{\bar o}}(b_{c,\alpha}) \left( {\mathbf 1}_{{\bar o}{\bar o}} 
-i {\mathbf A}_{{\bar o}{\bar o}}(b_{c,\alpha}) \right)^{-1} {\mathbf A}_{{\bar o} o}(b_{c,\alpha})
 - \dfrac{  B_{c,\alpha}^{\prime \prime}(b_{c,\alpha})  \bar{\mathbf u}_{o}(b_{c,\alpha}) {\mathbf C}(b_{c,\alpha})  \bar{\mathbf u}_{o}^t(b_{c,\alpha})}{2 (1 - B_{c,\alpha}^\prime(b_{c,\alpha}))^2 D_\alpha(b_{c,\alpha})}  \nonumber \\
&-& \dfrac{ (\bar{\mathbf u}_{o} {\mathbf C} \bar{\mathbf u}_{o}^t   )^\prime(b_{c,\alpha})}{  (1 - B_{c,\alpha}^\prime(b_{c,\alpha}))  D_\alpha(b_{c,\alpha})}  
+ \dfrac{ D_\alpha^\prime(b_{c,\alpha}) \bar{\mathbf u}_{o}(b_{c,\alpha}) {\mathbf C}(b_{c,\alpha})  \bar{\mathbf u}_{o}^t(b_{c,\alpha}) }{(1 - B_{c,\alpha}^\prime(b_{c,\alpha})) D_\alpha^2(b_{c,\alpha})  } 
- {\sum}_{\beta \neq \alpha} \dfrac{\mathbf{p}_\beta}{b_{c,\alpha}-b_{c,\beta}}
\eea
Several quantities need to be computed in order to evaluate the rhs of this equation.
The derivatives $B_{c,\alpha}^\prime$ and $B_{c,\alpha}^{\prime \prime}$ are obtained by first writing, for $B$ near $b_{c,\alpha}$,
${\mathbf B}_c(B) \approx {\mathbf B}_c(b_{c,\alpha}) + {\mathbf B}_c^\prime(b_{c,\alpha}) \eta$ with $\eta=( B - b_{c,\alpha})$ and the derivative ${\mathbf B}_c^\prime$ promptly computed from the definition of ${\mathbf B}_c$ in terms of ${\mathbf A}_{{\bar o}{\bar o}}$ and
of $\mathbf A^\prime_{{\bar o}{\bar o}}$.
Next, we apply perturbation theory with small parameter $\eta$ to the eigenvalue problem
${\mathbf B}_{c}(B) | \Omega_\alpha(B)\rangle =B_{c,\alpha}(B) | \Omega_\alpha(B) \rangle $
as
\be
\left( {\mathbf B}_c(b_{c,\alpha}) + {\mathbf B}_c^\prime(b_{c,\alpha}) \eta \right )
( | \Omega_\alpha^{(0)} +  | \Omega_\alpha^{(1)}\rangle \eta + \cdots )
=(b_{c,\alpha} +  b^{(1)}_{c,\alpha} \eta +  b^{(2)}_{c,\alpha} \eta^2 + \cdots) ( | \Omega_\alpha^{(0)} \rangle +  | \Omega^{(1)}_\alpha\rangle \eta +\cdots ),
\ee
where the superscript $^{(k)}$ denotes perturbation terms of order $k$ on eigenvalues and eigenfunctions. 
Equating $\eta$ powers order by order, one can compute terms up to perturbative order $k=2$ in the eigenvalues,
and identify $B_{c,\alpha}^\prime(b_{c,\alpha})=b^{(1)}_{c,\alpha}$ and $2 B_{c,\alpha}^{\prime \prime}(b_{c,\alpha})=b^{(2)}_{c,\alpha}$.

The only difference with the standard procedure is that matrix ${\mathbf B}_c$ is complex symmetric
rather than self-adjoint such that a generalization of perturbation theory to non-Hermitian operators 
must be employed in order to calculate $b^{(1)}_{c,\alpha}$ and $b^{(2)}_{c,\alpha}$.
It should be stressed that the only approximation involved in this procedure is the neglect of the second
derivative in the Taylor expansion of ${\mathbf B}_c(B)$ near $b_{c,\alpha}$.
This is justified since this term is usually small and it could in principle be included if needed,
even if at the expenses of significant added computational complexity to obtain $\mathbf A^{\prime \prime}_{{\bar o}{\bar o}}$.
Further, derivatives of the adjugate $\mathbf C$ are evaluated as explained below equation \eqref{eq_dAbgl}.
The terms $\bar{\mathbf u}_{o}^\prime$ are obtained based on the definition of $\bar{\mathbf u}_{o}$ 
and from the knowledge of $\mathbf A$ and ${\mathbf A}^\prime$.
Also note the useful formula 
\be
\dfrac{D^\prime_\alpha}{D_\alpha} = \log^\prime(D_\alpha) = {\sum}_{\beta\neq \alpha} 
\log^\prime(B-B_{c,\beta})   
= {\sum}_{\beta\neq \alpha} \dfrac{(1-B^\prime_{c,\beta})}{B-B_{c,\beta}}
\ee
allowing the ratio in last but one term on the rhs of \eqref{eq_abg_c} to be easily determined.

Once the pole strength ${\mathbf p}_\alpha$ known for each resonance, one can cast the 
energy-dependent complex scattering length \eqref{eq_c_pole} in the intuitive form
\be
\label{eq_intuitive}
{\mathbf a}(B) = {\mathbf a}_{\rm bg}(B) + \sum_{\alpha} \dfrac{{\mathbf p}_\alpha}{B-b_{c,\alpha}}
\ee
with
\be
{\mathbf a}_{\rm bg}(B) = - {\mathbf A}_{oo} -i {\mathbf A}_{o{\bar o}} \left( {\mathbf 1}_{{\bar o}{\bar o}} -i {\mathbf A}_{{\bar o}{\bar o}} \right)^{-1} {\mathbf A}_{{\bar o} o} - \bar{\mathbf u}_{o} ( B {\mathbf 1}_{Q} - {\mathbf B}_c)^{-1}   \bar{\mathbf u}_{o}^t   - \sum_{\alpha} \dfrac{{\mathbf p}_\alpha}{B-b_{c,\alpha}}
\ee
a quantity by construction regular at the poles. If $\mathbf A$ is strictly constant, it is
clear that last two terms exactly cancel and the background reduces to
\be
{\mathbf a}_{\rm bg}(B) = - {\mathbf A}_{oo} -i {\mathbf A}_{o{\bar o}} \left( {\mathbf 1}_{{\bar o}{\bar o}} -i {\mathbf A}_{{\bar o}{\bar o}} \right)^{-1} {\mathbf A}_{{\bar o} o}.
\ee

Generalizing the usual definition from the single-channel case \cite{Tiesinga1993}, we define the partial magnetic widths associated to pole $\alpha$ via the equation
\be
  {\mathbf \Delta}_{\alpha} = - {\mathbf a}_{\rm bg}^{-1/2}(b_{c,\alpha}) ~ {\mathbf p}_{\alpha}~  {\mathbf a}_{\rm bg}^{-1/2}(b_{c,\alpha}) .
\ee
In the absence of inelastic channels, the matrix ${\Delta}_{\alpha}$ is real, its imaginary part arises from inelastic processes.

As a final note, we have undertaken the approach to collect all degenerate channels in the same open-channel 
manifold, which led
us naturally to introduce matrices of scattering lengths and of magnetic widths. This description is most useful in
the presence of degenerate partial waves or in
equilibrium situations where all degenerate internal states are equally populated.
In other experiments one is more interested in polarizing the atoms in a single quantum state and in either monitoring the
evolution of the population or studying the collective behavior of the gaz in such specific state.
In this case, it will be more useful to identify the open channel manifold with the unique state of interest, and to
treat super-elastic channels on the same footing as the inelastic ones. All derivation above holds, the main difference being that the
complex scattering length will not be real even in the absence of inelastic channels, since the initial quantum
state will in general decay to degenerate channels. 

\begin{figure}[hb!]
\includegraphics[width=0.7\columnwidth] {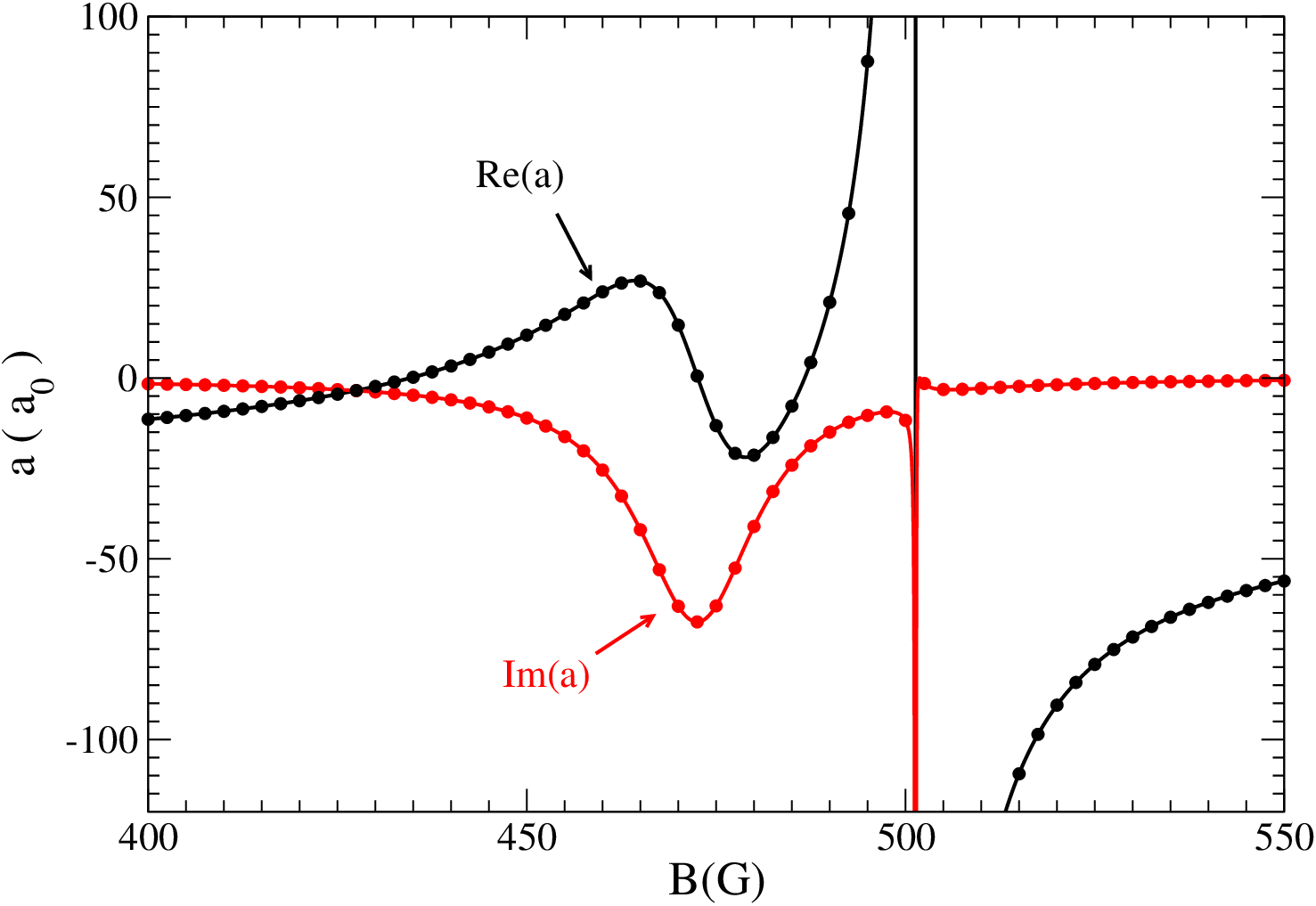}
\caption{
The real and the imaginary part of the complex scattering length
for collisions of two $^{39}$K atoms in hyperfine states $|1 ~ 1 \rangle$ and $|1~-1\rangle$.
Full curves represent results of a close-coupled numerical calculation, the dots 
the pole approximation of equation~\eqref{eq_intuitive} evaluated on a regular magnetic-field grid.
}
\label{fig1}
\end{figure}
\section{Comparison with other approaches}
\label{comparison}

The simplest case of an isolated elastic resonance in an elastic channel is trivially recovered from
equation \eqref{eq_intuitive} by dropping the $\alpha$ index, putting $p=-a_{\rm bg} \Delta$ and
realizing that the resonance center $b_c$ is a real quantity.

For the case of several overlapping resonances in an elastic channel,
simplicity of equation \eqref{eq_intuitive} may seem at first sight surprising as compared to
equation~(26) of Jachymski and Julienne~\cite{Jachymski2013}. 
In order to show that the present approach reduces to the quantum defect analysis
presented in the latter work, we give a brief alternative derivation of their result
starting from the present standard scattering theory. 
To this aim, let us specialize \eqref{eq_kmat_Fesh} to the case 
of a single open channel and consider by the sake of simplicity the low-energy 
limit :
\be
{\mathbf K} =
 \cos({ \xi_{\rm bg}}) \left[ \tan({\xi}_{\rm bg}) + {\mathbf
K}_{\rm res}  \right]  \left[ {\bm 1}_P - \tan({ \xi}_{\rm bg}) {\mathbf K}_{\rm
res}   \right]^{-1} \cos({ \xi_{\rm bg}})^{-1} 
\underset{E \to 0}{\to} 
 \tan({\xi}_{\rm bg}) + {\mathbf
K}_{\rm res}  
\ee

In the approach of Jachymski and Julienne, one first 
factors the energy dependence out of the predissociation amplitude ${\mathbf y}$ 
using the energy-dependent quantum defect amplitude $C$ as ${\mathbf y}=C^{-1} \bar{\mathbf y}$.
Then one recognizes that the shift matrix in the denominator of ${\mathbf K}_{\rm
res}$ contains an energy-dependent part
proportional to $\bar{\mathbf y} \bar{\mathbf y}^t$ via the quantum defect function 
$\tan( \lambda )$ \cite{Bohn1999}. This leads to :
\be
\label{eq_jul_a}
{\mathbf K} =  \tan({\xi}_{\rm bg}) + C^{-2} \bar{\mathbf y}^t
\left( {\mathbf D - \tan( \lambda ) \bar{\mathbf y} \bar{\mathbf y}^t} \right)^{-1}
\bar{\mathbf y}  \quad , \quad  E \to 0
\ee
where $\mathbf D$ contains the Zeeman shift $-{\bm \mu} \left( B {\mathbf 1}_Q - {\mathbf B}_r \right)$ 
and all energy-independent couplings between channels
in $Q$ space. 
Making use of Woodbory's matrix identity to express the inverse, equation \eqref{eq_jul_a}
is seen to be equivalent to 
\be
\label{eq_jul_b}
{\mathbf K} =  \tan({\xi}_{\rm bg}) + C^{-2} 
\frac{ \bar{\mathbf y}^t  {\mathbf D}^{-1} \bar{\mathbf y} }{ 1 - \tan( \lambda ) \bar{\mathbf y}^t {\mathbf D}^{-1} \bar{\mathbf y}}   \quad , \quad  E \to 0
\ee
Assuming absence of coupling between molecular states near threshold, matrix
$\mathbf D$  takes 
the simple diagonal form 
${\mathbf D} =-{\bm \mu} \left( B {\mathbf 1}_Q - {\mathbf B}_r \right)$. 
Replacing in \eqref{eq_jul_b} leads after simple manipulations to
equation~(26) of~\cite{Jachymski2013}. 

Note that in the present work one first diagonalizes the denominator
of ${\mathbf K}_{\rm res}$ at each value of the magnetic field.
Therefore, our equation \eqref{eq_intuitive} can be considered in a sense the adiabatic version of
equation~(26) of Jachymski and Julienne. Note however that equation~\eqref{eq_intuitive} in the present work has
a somewhat broader domain of applicability, since it does not rely on the assumption that the molecular states
are uncoupled near threshold.

We now consider an inelastic isolated resonance. With reference to the notation of equation (1) in \cite{Kempen2002} their lineshsape is recovered
by first writing our complex pole $p$ in the form $p = -\Delta a_{\rm bg}$, then by expressing the complex width $\Delta$ in polar form
as $ e^{2 i \Phi_R} \Delta_{\rm el}$. Finally, one identifies $a_{\rm bg}$ with $a_\infty$ and the complex magnetic field location $b_c$ in 
the denominator of \eqref{eq_intuitive} with the term $B_0 -i \dfrac{\Delta_{\rm inel}}{2}$ in \cite{Kempen2002}. 

Barring from the choice of the notation, it is also easy to verify the equivalence of the lineshapes in \cite{Hutson2007} and
\cite{Kempen2002} and thus in the present work.

\section{Applications}
\label{applications}
\subsection{Overlapping inelastic $s$-wave resonances}
As a first application, let us consider collisions of ultracold $^{39}$K atoms, a system whose 
scattering properties
have been determined very precisely first based on photoassociation then
on magnetic Feshbach resonance spectroscopy \cite{2007-CDE-NJP-223}.
We consider here the case of $s$-wave collisions
for atoms in Zeeman sublevels correlating in zero magnetic field
with hyperfine quantum numbers $ ( F=1,~ M=1 )$ and $ ( F=1,~ M=-1 )$, respectively.
This combination is not stable under spin-exchange collisions, since it can decay to
a pair of $ ( F=1,~ M=0 )$ atoms.

\begin{figure}[ht]
\includegraphics[width=0.6\columnwidth] {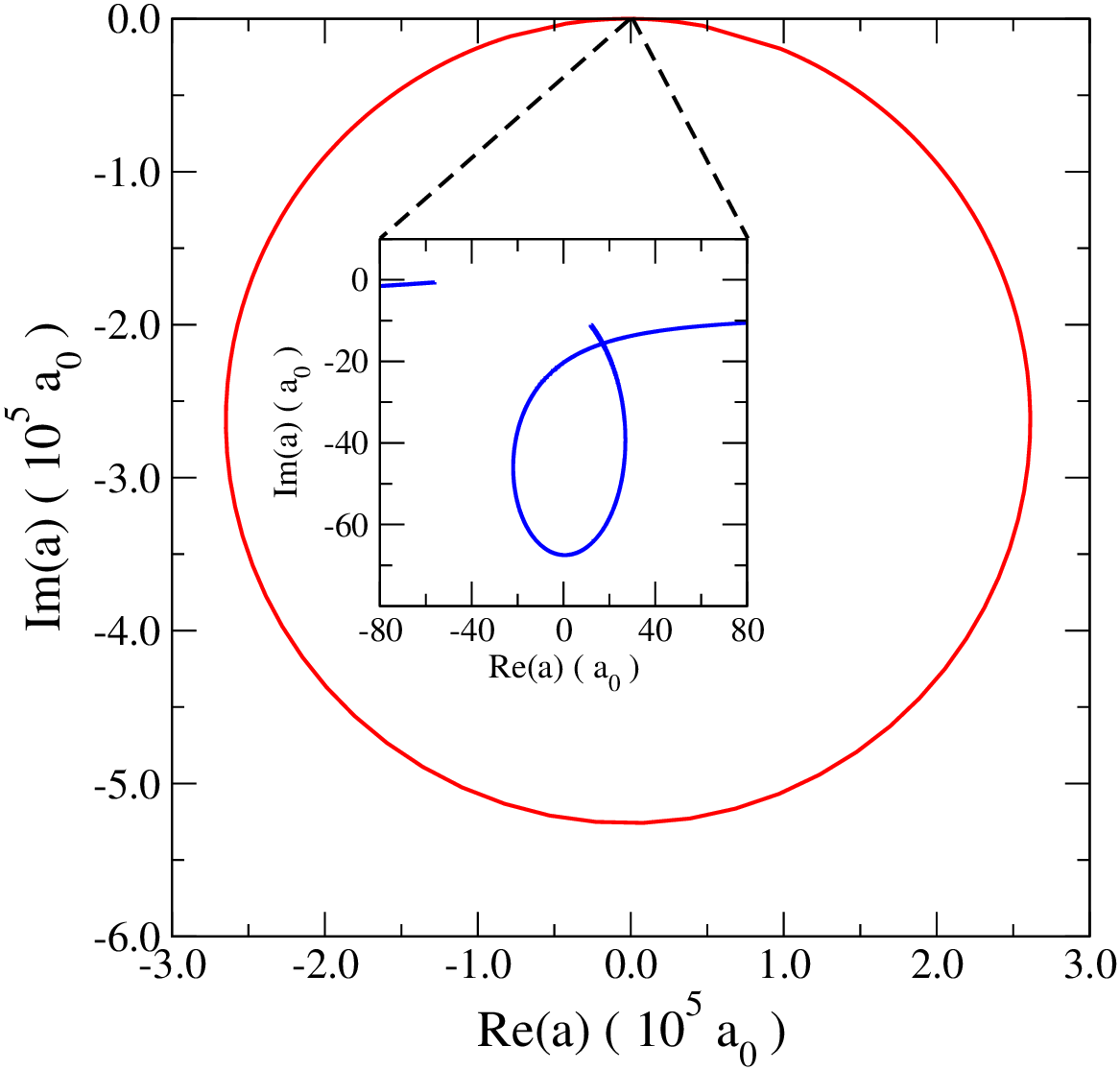}
\caption{
Trajectory in the complex plane of the complex scattering length when the magnetic field spans
the features of figure \ref{fig1}. The inset shows a smaller loop described near the origin
by $a(B)$ as the magnetic field varies across the weaker of the two features. Note the largely different
scales on the axes of the main figure and the inset.
}
\label{fig2}
\end{figure}

The complex scattering length shows two features, which have been reported before~\cite{Lysebo2008}.
Based on the much stronger variation of the real part of $a$ near 475~G than near 500~G, the authors of \cite{Lysebo2008}
concluded that only one feature is of resonant origin. 
A closer analysis of figure \ref{fig1} shows however that the real part of $a$ does not diverge near either 
feature, as expected on a general ground in the presence of inelastic processes. 
While the feature at higher field is indeed significantly more pronounced, one may still suspect the 
presence of two quasi-bound states underlying the observed structure.
\begin{figure}[ht!]
\includegraphics[width=0.7\columnwidth] {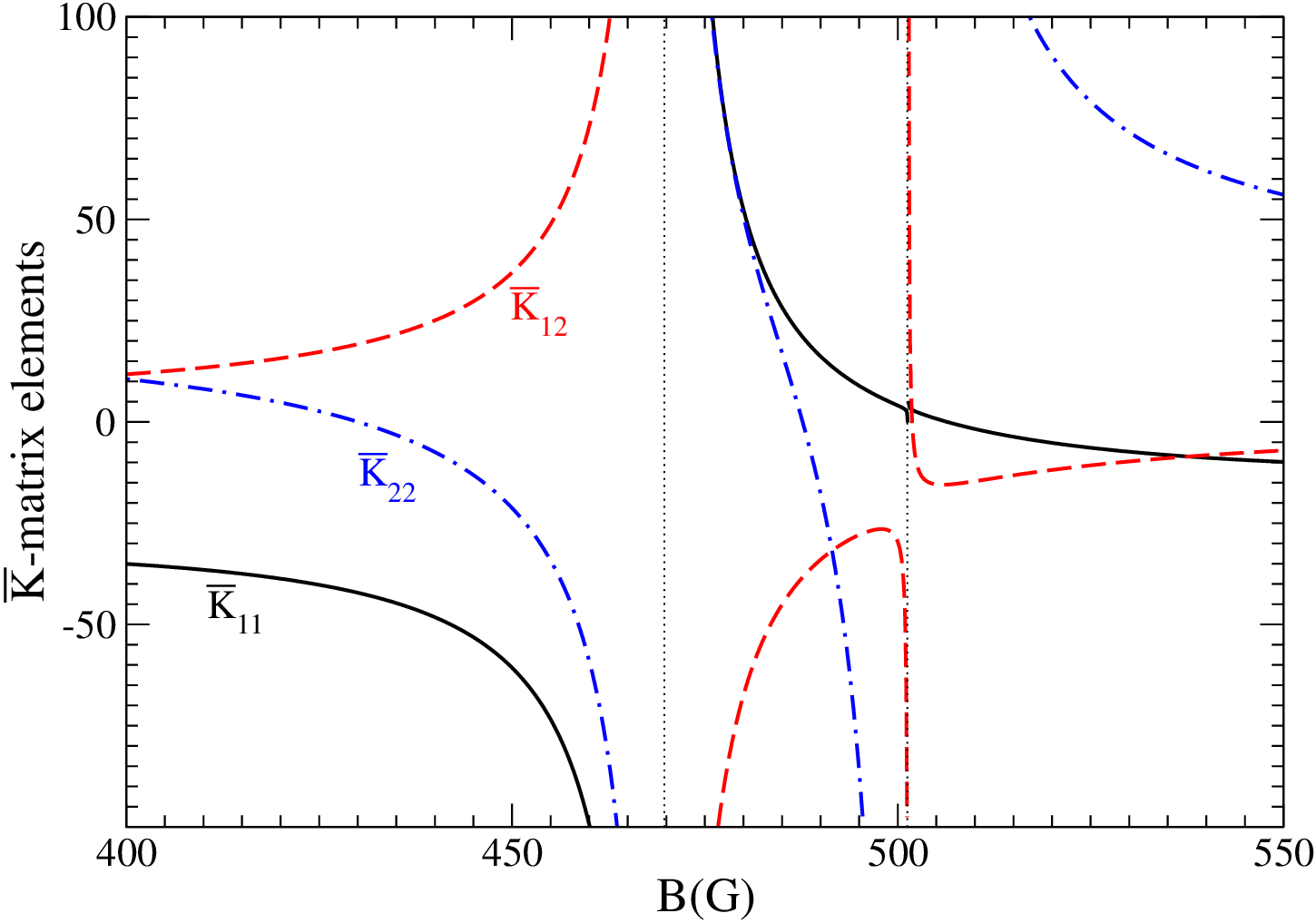}
\caption{
The elements of the normalized matrix $\bar{\mathbf K}$, namely
$\bar{\mathbf K}_{11}$ (full line), $\bar{\mathbf K}_{12}$ (dashed
line) and $\bar{\mathbf K}_{22}$ (dash-dotted line) for for collisions
of $^{39}$K atoms in hyperfine states $|1~1\rangle$ and $|1~-1\rangle$.
Vertical dotted lines indicate the location of two neighboring poles.
Indices $1$ and $2$ denote the two-body collisions channels $| 1~ 0 \rangle \otimes |1 0 \rangle
$ and $| 1~1 \rangle \otimes | 1~ -1 \rangle$, respectively.
}
\label{fig3}
\end{figure}

A complimentary picture is provided by the trajectory of $a(B)$ in the complex scattering-length plane as a function
of the real parameter $B$; See figure \ref{fig2}. It it easy to verify that in the ideal case of an isolated resonance with strictly
constant background such trajectory is a circle \cite{Hutson2007}. One can actually observe in the figure one large circle
associated with the strong resonance, whereas closer inspection near the origin (see inset) highlights the presence of a second
deformed loop described as the magnetic field crosses the weak feature, suggesting again the presence of a resonance pair. 
One can also observe that the global trajectory does not exactly close, since the values of the background
scattering length away from resonance are not identical on the left and right resonance sides.

Conclusive evidence about the nature of this feature is obtained by inspecting the elements of the full 
$K$-matrix depicted in figure \ref{fig3}. 
In agreement with equation \eqref{kmatgenB_b}, all elements of the reactance matrix diverge on resonance at the same location.
Two poles are clearly observed, confirming that the relevant $Q$ subspace contains two bare bound states coupled to the incoming channel.
Following the numerical procedure outlined in section \ref{formalism}, it is now easy to extract well defined
lineshape parameters without resorting to a fit. Resulting numerical values for the complex background scattering length, field 
location and magnetic width can be found in table \ref{tab_1}.

It is also interesting to compare graphically the exact numerical result with the pole approximation \eqref{eq_intuitive}
represented by the dots in figure \ref{fig1}. 
Note from the numerical values in the table that the background is nearly constant and can be represented locally by an interpolating linear function. 
As expected, the agreement with the non trivial behavior of the full numerical calculation is excellent, confirming that 
all elements of the theory have been 
computed and integrated properly.

\begin{table}[bh!]
\begin{center}
\caption{
Complex lineshape parameters (see text for definitions) appearing in lineshape \eqref{eq_intuitive}, with powers of tens indicated in brackets. 
Sample resonance parameters for ultracold collisions of K atoms in hyperfine states and partial wave specified in the first column are reported. For comparison, calculations with and without the spin-spin interaction are reported for $p$-wave scattering.
}
\label{tab_1}
\vskip 12pt
\begin{tabular}{|l |   c c c c  |}
\hline 
{\rm Channel}                    & $a_{\rm bg} (a_0^{ \ell+1})$  &    $b_c$(G) & $p_\alpha (a_0^{\ell+1} {\rm G})$  &  $\Delta_\alpha$(G)   \\
\hline 
K($|1 ~1\rangle$)+K($| 1 ~-1 \rangle$) $s$-wave           &  $ 29.1 +  5.37[-2] i$          &  $472.67 + 9.45 i$         & $-639+26.9 i$ &   $21.9- 0.965 i$                   \\
                                 &  $ 29.3 +  4.97[-2] i  $       &  $501.34 + 1.70[-3] i $        & $-897+5.91 i$ &   $30.6 - 0.253 i  $              \\
K($|1 ~1\rangle$)+K($| 1 ~0 \rangle$) $p$-wave ~ {\rm (with spin-spin)}          &  $ -9.28[4] $          &  $379.1$           &$-4.00[6]$  &   $ -43.1 $                   \\
                                 &  $ -9.60[4] $          &  $400.7  $               & $ -1.27[6] $  &   $ -1.32 $                   \\
                                 &  $ -1.16[5] $          &  $720.3  $               & $ -3.80[2] $ &   $ -3.26[-3]  $                   \\
                                 &  $ -1.17[5] $          &  $757.8  $               & $ -2.38[3] $ &   $ -2.03[-2] $                   \\
                                 &  $ -1.17[5] $          &  $762.3  $               & $ -3.20[4] $  &   $ -0.272 $                   \\
K($| 1 ~1 \rangle$)+K($| 1 ~0 \rangle$) $p$-wave ~ {\rm (w/o spin-spin)}          &  $2.81[5]  $          &  $379.2  $    &  $-4.01[6] $         &     $ 14.3 $                   \\
                                                      &  $2.81[5]  $          &  $400.6  $    &  $-1.27[6] $        &     $ 0.451 $                   \\
                                                      &  $2.87[5]  $          &  $758.0  $    &  $-2.38[3] $         &     $ 8.30[-3] $                   \\
                                                      &  $2.87[5]  $          &  $762.6  $    &  $-3.21[4] $         &     $ 0.111 $                   \\
\hline
\end{tabular}
\end{center}
\end{table}

\subsection{Energy-dependent $p$-wave scattering volume}

As a further example, let us consider collisions in higher-order partial waves.
Specifically, we consider $p$-wave collisions between a pair of $^{39}$K atoms 
prepared respectively in states $|1~1\rangle$ and $|1~0\rangle$ at a finite collision energy of 
$E / k_{\rm B} = 1 \mu {\rm K}$.
We consider the projection $M=2$ of the total axial angular momentum, corresponding to
a pair of atoms with initial axial mechanical angular momentum $m=1$. In this symmetry block, if $p$ waves only are
included, scattering is purely elastic.
The energy-dependent scattering volume is defined according to the standard definition \eqref{eq_asc}.
\begin{figure}[ht]
\includegraphics[width=0.7\columnwidth] {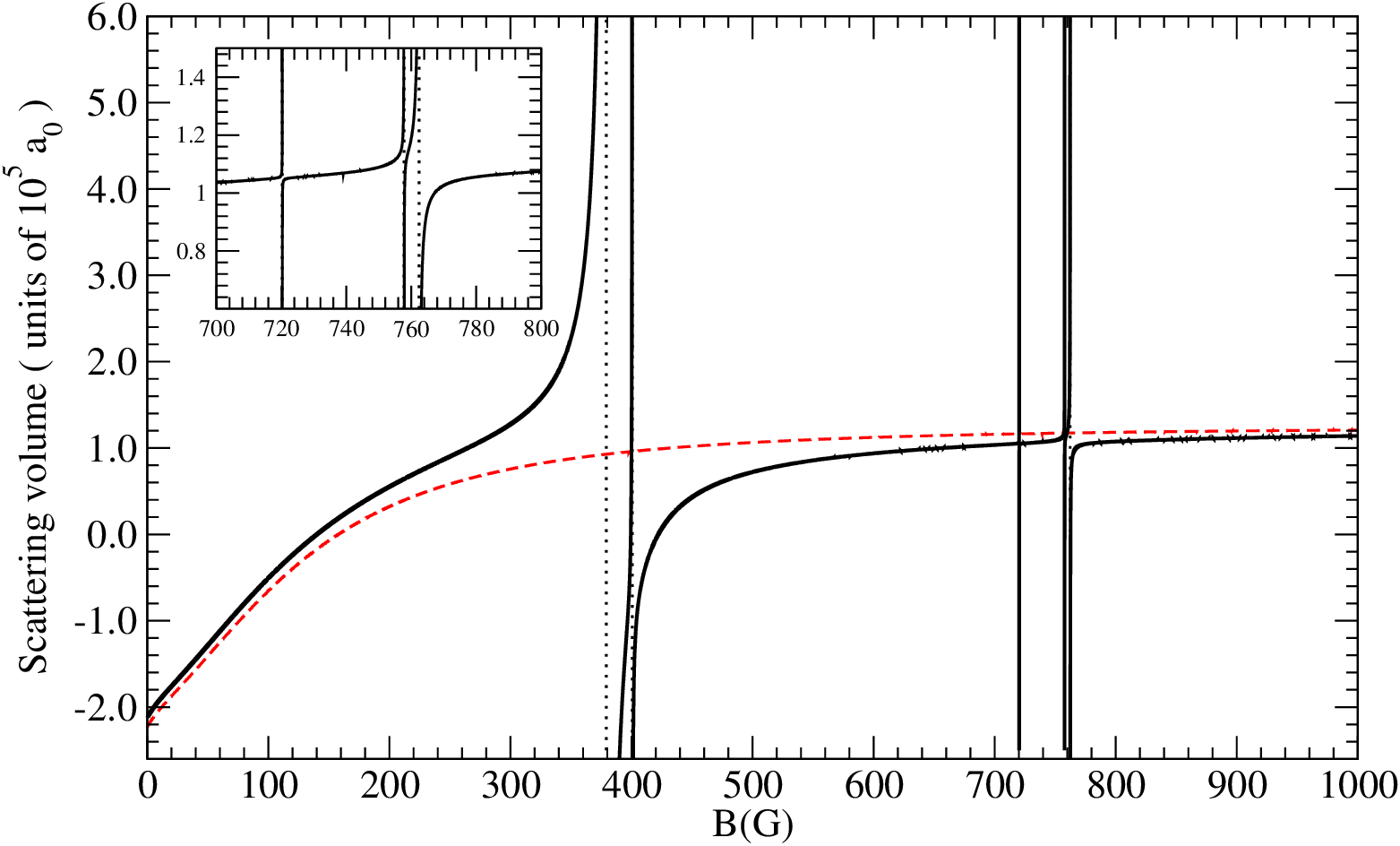}
\caption{
Energy dependent scattering volume for collisions of two $^{39}$K atoms in hyperfine states $| 1 1 \rangle$ and $|1 0 \rangle$ (full line).
The dashed line represents the background scattering volume (see text). The inset shows the detail of a triplet of resonances
above 700~G, of which two overlapping.
}
\label{fig4}
\end{figure}

\begin{figure}[ht]
\includegraphics[width=0.7\columnwidth] {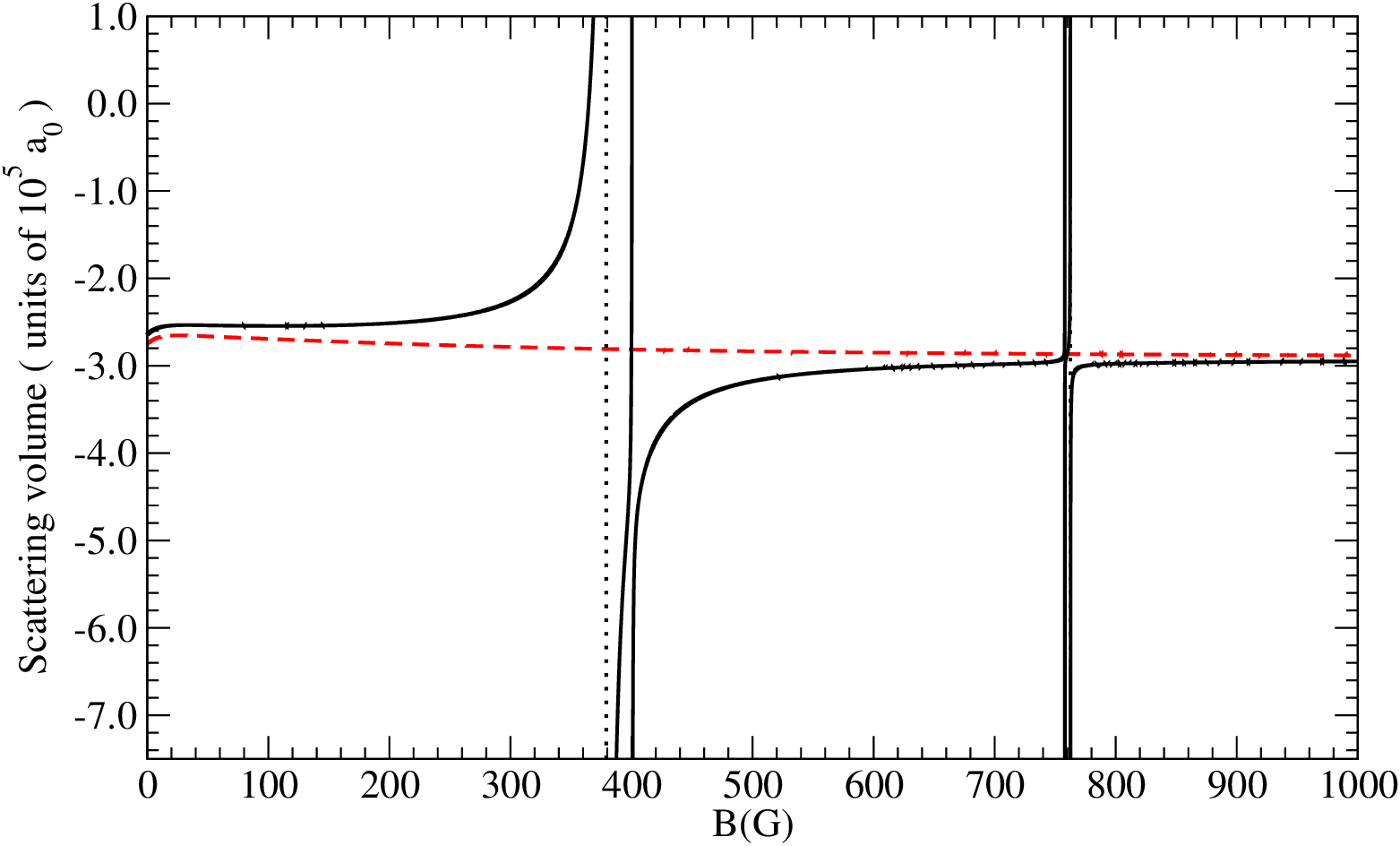}
\caption{
Same as figure \ref{fig4}, but with the spin-spin interaction set artificially to zero.
}
\label{fig5}
\end{figure}

One can observe in figure \ref{fig3} two series of overlapping resonances in the range of magnetic fields below $1000$G, 
a doublet near 400~G and a triplet near 750~G. The largest resonance at 379~G is sufficiently broad to 
modify the local background near 750~G.
Visually, the background appears to vary in a non trivial way with the magnetic field, in particular near the 
origin. 
The present approach has the advantage that no modeling is required in order to extract the resonance parameters.
It is sufficient to locate precisely the resonances, extract the corresponding pole residues, and finally the
local and global background at the resonance positions.
The corresponding parameters for the present case are reproduced in table \ref{tab_1}.

If needed, the field-dependent background scattering length can be extracted by subtracting the pole contribution
from the full scattering volume, {\it i.e.} last term on the rhs of \eqref{eq_intuitive} from the lhs. 
In spite of the fact that such operation involves taking the difference of large numbers and as such is 
in principle numerically flawed very close to the poles, we obtain a smooth background that, as anticipated, varies 
relatively fast in a non trivial way from negative to positive values.

Without performing a detailed analysis that is not the aim of the present methodological work, it is interesting to stress
the effect of the spin-spin dipolar interaction. 
To this extent we set artificially to zero the dipolar coupling constant and perform
the same steps as above. In the absence of anisotropic interactions, not only $M$ but also $M_f = M_{a} + M_{b}$
is a good quantum number. This implies that resonances associated with molecular states with $M_f$ different from the
one of the incoming atoms are completely uncoupled, {\it i.e.} have zero width.
Indeed, one can remark in figure \ref{fig4} the absence of very narrow resonances observed in figure \ref{fig3}.
The remaining relatively broad resonances occur in blocks with given $M_f$ and are induced by the interplay of spin-exchange and spin-spin
interactions.
In the present case, spin-exchange is found to largely dominate over dipolar interactions, as demonstrated by the fact that the pole strength $p$
in the presence and in the absence of spin-spin couplings (see table) differ by no more less than $1\%$. Similarly, the
spin-spin interaction only weakly (below 300~mG) shifts the resonance position.
On the other side, the background scattering volume (and {\it a-fortiori} the magnetic width), has a significantly stronger variation with magnetic field in the presence 
of the dipolar interaction, highlighting the strong influence of the spin-spin interaction on the incoming wavefunction.

\section{Conclusions}
\label{conclusions}
We have presented a formalism allowing resonance parameters to be extracted from a
numerical calculation in a well determined algorithmic way. 
In our opinion the present approach fully solves a longstanding difficulty in the analysis of 
resonance spectra involving multiple overlapping features or a non trivial behavior of background scattering.
Our results also provide a simple and intuitive lineshape for inelastic overlapping resonances that has been demonstrated
starting from first principles. 
The machinery presented in this work can be incorporated in existing computer packages, provided that the underlying algorithm
to solve the Schr\"odinger equation allows the wavefunction derivatives to be easily evaluated.
As a complimentary step to the present paper that focuses on the lineshape parameters, it could
be interesting to analyze the wavefunction at the poles of the reactance matrix in order to gain information on the nature
of the resonant states.

\begin{acknowledgments}
Useful exchanges with P. S. Julienne and K. Jachymski are gratefully acknowledged.
\end{acknowledgments}

\appendix
\section{Proof of various equations}
\label{appendix_1}
Proof of equation~\eqref{kmatgenB} goes as follows. Substituting
\eqref{kresredef} into \eqref{eq_kmat_Fesh}, using Woodbury matrix identity and the definition of $\hat{\bm {\mathcal B}}$, one gets
\bea
{\mathbf K} &=& 
\cos({\bm \xi_{\rm bg}}) \left[ \tan({\bm \xi}_{\rm bg}) + 
\tilde{\mathbf y} \left( B{\bm 1}_Q - {\bm {\mathcal B}}  \right)^{-1} 
\tilde{\mathbf y}^t
 \right]  \left[ {\bm 1}_P - \tan({\bm \xi}_{\rm bg}) 
\tilde{\mathbf y} \left( B{\bm 1}_Q - {\bm {\mathcal B}}  \right)^{-1} \tilde{\mathbf y}^t
\right]^{-1} \cos({\bm \xi_{\rm bg}})^{-1} 
\nonumber \\
&=&\cos({\bm \xi_{\rm bg}}) \left[ \tan({\bm \xi}_{\rm bg}) +
\tilde{\mathbf y} \left( B{\bm 1}_Q - {\bm {\mathcal B}}  \right)^{-1}
\tilde{\mathbf y}^t
 \right] 
 \left[ {\bm 1}_P + \tan({\bm \xi}_{\rm bg})
\tilde{\mathbf y} \left( B{\bm 1}_Q - {\bm {\mathcal B}} 
-  \tilde{\mathbf y}^t \tan({\bm \xi}_{\rm bg}) \tilde{\mathbf y}  \right)^{-1} \tilde{\mathbf y}^t
\right] \cos({\bm \xi_{\rm bg}})^{-1}
\nonumber \\
&=&\cos({\bm \xi_{\rm bg}}) \left[ \tan({\bm \xi}_{\rm bg}) +
\tilde{\mathbf y} \left( B{\bm 1}_Q - {\bm {\mathcal B}}  \right)^{-1}
\tilde{\mathbf y}^t
 \right]
 \left[ {\bm 1}_P + \tan({\bm \xi}_{\rm bg})
\tilde{\mathbf y} \left( B{\bm 1}_Q - \hat{\bm {\mathcal B}} 
 \right)^{-1} \tilde{\mathbf y}^t
\right] \cos({\bm \xi_{\rm bg}})^{-1}
\eea
Carrying out the inner product, replacing the term $ \tilde{\mathbf y}^t \tan({\bm \xi}_{\rm bg}) \tilde{\mathbf y} $ by $(\hat{\bm {\mathcal B}}
- {\bm {\mathcal B}})$ in the numerator and factoring by grouping
\bea 
\label{eq_app_1}
{\mathbf K} &=&
\cos({\bm \xi_{\rm bg}}) \left\{  \tan({\bm \xi}_{\rm bg}) + \tilde{\mathbf y} 
\left[ ( B{\bm 1}_Q - {\bm {\mathcal B}})^{-1} + ( B{\bm 1}_Q - {\bm {\mathcal B}})^{-1}
(\hat{\bm {\mathcal B}}
- {\bm {\mathcal B}})   ( B{\bm 1}_Q - \hat{\bm {\mathcal B}})^{-1}
      \right]      \tilde{\mathbf y}^t      
\right\} \cos({\bm \xi_{\rm bg}})^{-1}  \nonumber \\
&+& \cos({\bm \xi_{\rm bg}}) \tan^2({\bm \xi_{\rm bg}})  \tilde{\mathbf y} ( B{\bm 1}_Q - \hat{\bm {\mathcal B}})^{-1} \tilde{\mathbf y}^t .
\eea 
Finally noticing that
\be
\label{eq_simple_mat_alg}
  \left[ ( B{\bm 1}_Q - {\bm {\mathcal B}})^{-1} + ( B{\bm 1}_Q - {\bm {\mathcal B}})^{-1}
(\hat{\bm {\mathcal B}}
- {\bm {\mathcal B}})   ( B{\bm 1}_Q - \hat{\bm {\mathcal B}})^{-1}
      \right] = ( B{\bm 1}_Q - {\bm {\mathcal B}})^{-1}  
\left[  B{\bm 1}_Q -  \hat{\bm {\mathcal B}} +  \hat{\bm {\mathcal B}} 
-   {\bm {\mathcal B}} \right] ( B{\bm 1}_Q - \hat{\bm {\mathcal B}})^{-1}
= B{\bm 1}_Q - \hat{\bm {\mathcal B}} ,
\ee
substituting in \eqref{eq_app_1} and using simple trigonometric identities,
equation~\eqref{kmatgenB} easily follows.

To prove \eqref{eq_c_pole}, we start by expanding \eqref{eq_int} as
\bea
\label{eq_int_app}
{\mathbf a}(B)  &=&
 - {\mathbf A}_{oo}  -i {\mathbf A}_{o \bar o}   (  {\mathbf 1}_{\bar o \bar o} -i {\mathbf A}_{\bar o \bar o}  )^{-1}  {\mathbf A}_{\bar o o}  
+ {\mathbf A}_{o \bar o} {\mathbf u}_{\bar o} ( B {\mathbf 1}_{Q} - {\mathbf B}_c)^{-1}  {\mathbf u}^t_{\bar o} 
{\mathbf A}_{{\bar o} o} - \bar{\mathbf y}_{o}  ( B {\mathbf 1}_Q - {\mathbf b})^{-1} \bar{\mathbf y}^t_{\bar o}
\nonumber \\
&+&  \bar{\mathbf y}_{o} ( B {\mathbf 1}_Q - {\mathbf b})^{-1}  \bar{\mathbf y}_{\bar o}^t  {\mathbf u}_{\bar o}
( B {\mathbf 1}_{Q} - {\mathbf B}_c)^{-1} {\mathbf u}_{\bar o}^t  \bar{\mathbf y}_{\bar o}  ( B {\mathbf 1}_Q - {\mathbf b})^{-1}  \bar{\mathbf y}_{o}^t  
- i \bar{\mathbf y}_{o} ( B {\mathbf 1}_Q - {\mathbf b})^{-1}  \bar{\mathbf y}_{\bar o}^t  
 (  {\mathbf 1}_{\bar o \bar o} -i {\mathbf A}_{\bar o \bar o}  )^{-1}  \bar{\mathbf y}_{\bar o}  ( B {\mathbf 1}_Q - {\mathbf b})^{-1}  \bar{\mathbf y}_{o}^t
\nonumber \\
&-& i {\mathbf A}_{o \bar o}  (  {\mathbf 1}_{\bar o \bar o} -i {\mathbf A}_{\bar o \bar o}  )^{-1} 
 \bar{\mathbf y}_{\bar o}  ( B {\mathbf 1}_Q - {\mathbf b})^{-1}  \bar{\mathbf y}_{o}^t
+  {\mathbf A}_{o \bar o}  {\mathbf u}_{\bar o} ( B {\mathbf 1}_{Q} - {\mathbf B}_c)^{-1}  {\mathbf u}^t_{\bar o}
\bar{\mathbf y}_{\bar o}  ( B {\mathbf 1}_Q - {\mathbf b})^{-1} \bar{\mathbf y}^t_{o}
\nonumber \\
&-&i \bar{\mathbf y}_{o} ( B {\mathbf 1}_Q - {\mathbf b})^{-1} \bar{\mathbf y}_{\bar o}^t 
 (  {\mathbf 1}_{\bar o \bar o} -i {\mathbf A}_{\bar o \bar o}  )^{-1} {\mathbf A}_{\bar o o}
+ {\mathbf A}_{o \bar o}  {\mathbf u}_{\bar o}   ( B {\mathbf 1}_Q - {\mathbf b})^{-1} {\mathbf u}_{\bar o}^t
 \bar{\mathbf y}_{\bar o} ( B {\mathbf 1}_Q - {\mathbf B}_c)^{-1} \bar{\mathbf y}_{o}^t  .
\eea

Let us now transform second line. Using definitions
of ${\mathbf B}_c$ and ${\mathbf u}_{\bar o}$, one can replace the terms
$\bar{\mathbf y}_{\bar o}^t  {\mathbf u}_{\bar o}$, ${\mathbf u}_{\bar o}^t  \bar{\mathbf y}_{\bar o}  
$, and $\bar{\mathbf y}_{\bar o}^t
 (  {\mathbf 1}_{\bar o \bar o} -i {\mathbf A}_{\bar o \bar o}  )^{-1}  \bar{\mathbf y}_{\bar o}$ by $ ({\mathbf B}_c -   {\mathbf b} ) / i$.
Regrouping and simplifying, second line transforms into :
\bea
\label{eq_second_line}
&-&\bar{\mathbf y}_{o} ( B {\mathbf 1}_Q - {\mathbf b})^{-1} 
\left[  ({\mathbf B}_c -  {\mathbf b}) ( B {\mathbf 1}_Q - {\mathbf B}_c  )^{-1} ({\mathbf B}_c -   {\mathbf b} ) + ({\mathbf B}_c -   {\mathbf b}  )     \right ]
 ( B {\mathbf 1}_Q - {\mathbf b})^{-1} \bar{\mathbf y}^t_{o} \nonumber \\
=&-& \bar{\mathbf y}_{o} ( B {\mathbf 1}_Q - {\mathbf b})^{-1}   ( B {\mathbf 1}_Q - {\mathbf B}_c)^{-1}  ( B {\mathbf 1}_Q - {\mathbf b})^{-1}
\bar{\mathbf y}^t_{o}.
\eea
Adding last term in first line of \eqref{eq_int_app} to result~\eqref{eq_second_line} yields after simple
matrix algebra similar to the manipulations in \eqref{eq_simple_mat_alg}
\bea
\label{eq_second_line_a}
 &-&\bar{\mathbf y}_{o} ( B {\mathbf 1}_Q - {\mathbf b})^{-1}   ( B {\mathbf 1}_Q - {\mathbf B}_c)^{-1}  ( B {\mathbf 1}_Q - {\mathbf b})^{-1}
\bar{\mathbf y}^t_{o}  - \bar{\mathbf y}_{o} \left( B {\mathbf 1}_Q - {\mathbf b} \right)^{-1} \bar{\mathbf y}^t_{o}
\nonumber \\
= &-& \bar{\mathbf y}_{o} ( B {\mathbf 1}_Q - {\mathbf B}_c)^{-1}  \bar{\mathbf y}^t_{o}
\eea
Along the same lines, one can simplify third
and last line in~\eqref{eq_int_app}. For instance, for the third line one has
\bea
\label{eq_second_but_last_line}
&-&i {\mathbf A}_{o \bar o}  (  {\mathbf 1}_{\bar o \bar o} -i {\mathbf A}_{\bar o \bar o}  )^{-1}
 \bar{\mathbf y}_{\bar o}  ( B {\mathbf 1}_Q - {\mathbf b})^{-1}  \bar{\mathbf y}_{o}^t
+  {\mathbf A}_{o \bar o}  {\mathbf u}_{\bar o} ( B {\mathbf 1}_{Q} - {\mathbf B}_c)^{-1}  {\mathbf u}^t_{\bar o}
\bar{\mathbf y}_{\bar o}  ( B {\mathbf 1}_Q - {\mathbf b})^{-1} \bar{\mathbf y}^t_{o} 
\nonumber \\
= &-&i {\mathbf A}_{o \bar o}  {\mathbf u}_{\bar o}
\left[  
( B {\mathbf 1}_Q - {\mathbf b})^{-1} + ( B {\mathbf 1}_Q - {\mathbf B}_c)^{-1}
(  {\mathbf B}_c - {\mathbf b})  ( B {\mathbf 1}_Q - {\mathbf b})^{-1}
\right ] \bar{\mathbf y}^t_{o}
\nonumber \\
= &-& i {\mathbf A}_{o \bar o}   {\mathbf u}_{\bar o}
( B {\mathbf 1}_{Q} - {\mathbf B}_c)^{-1} \bar{\mathbf y}^t_{o}
\eea
where first equality follows upon using the definition of ${\mathbf u}_{\bar o}$ and 
using again $\bar{\mathbf y}_{\bar o}^t  {\mathbf u}_{\bar o}=-i ({\mathbf B}_c -   {\mathbf b} )$, and collecting common terms.
Second equality results from simple matrix algebra.
Finally, last line in~\eqref{eq_int_app} can be expressed as 
\bea 
\label{eq_last_line}
&-&i \bar{\mathbf y}_{o} ( B {\mathbf 1}_Q - {\mathbf b})^{-1} \bar{\mathbf y}_{\bar o}^t
 (  {\mathbf 1}_{\bar o \bar o} -i {\mathbf A}_{\bar o \bar o}  )^{-1} {\mathbf A}_{\bar o o}
+ {\mathbf A}_{o \bar o}  {\mathbf u}_{\bar o}   ( B {\mathbf 1}_Q - {\mathbf b})^{-1} {\mathbf u}_{\bar o}^t
 \bar{\mathbf y}_{\bar o} ( B {\mathbf 1}_Q - {\mathbf B}_c)^{-1} \bar{\mathbf y}_{o}^t   \nonumber \\
=&-& i \bar{\mathbf y}_{o} 
( B {\mathbf 1}_{Q} - {\mathbf B}_c)^{-1}
{\mathbf u}_{\bar o}^t {\mathbf A}_{{\bar o} o}
\eea
Substituting equations~\eqref{eq_second_line},~\eqref{eq_second_but_last_line} and~\eqref{eq_last_line} into~\eqref{eq_int_app}, collecting terms with common factor
$( B {\mathbf 1}_Q - {\mathbf B}_c)^{-1}$ and defining, according to the main text,
$\bar {\mathbf u}_{ o}= i {\mathbf A}_{o{\bar o}} {\mathbf u}_{\bar o} + \bar{\mathbf y}_{ o}$
, \eqref{eq_c_pole} promptly follows.

\section*{References}

\end{document}